\documentclass[journal]{IEEEtran}
\ifCLASSINFOpdf
\else
\fi
\usepackage{tikz}
\newcommand*{\QEDB}{\hfill\ensuremath{\square}}%
\def\ci{\perp\!\!\!\perp}
\usepackage{amsmath,amssymb}
\usepackage{makeidx}
\usepackage{amssymb}
\usepackage{graphicx}
\usepackage{color}
\newtheorem{teiri}{Theorem}
\newtheorem{prop}{Proposition}
\newtheorem{rei}{Example}
\newtheorem{hodai}{Lemma}
\begin{document}
%
\title{Forest Learning from Data\ and its Universal Coding}
%
%
%

\author{Joe~Suzuki~\IEEEmembership{}
\thanks{
Department of Mathematical Science, Graduate School of Engineering Science, Osaka University,
Toyonaka, Osaka 560-8531, Japan, 
e-mail: j-suzuki@sigmath.es.osaka-u.ac.jp}
\thanks{Manuscript received April 19, 2016; revised July 5, 2018. This paper was partially presented at IEEE
International Symposium on Information Theory, Barcelona, Spain, July 2016.}
}

%
%

\markboth{IEEE TRANSACTIONS ON INFORMATION THEORY,~Vol.~61, No.~12, DECEMBER~2018}{JOE SUZUKI: FOREST LEARNING and UNIVERSAL CODING}
%



\maketitle

\begin{abstract}
This paper considers structure learning from data with $n$ samples of $p$ variables, assuming that 
the structure is a forest, using the Chow-Liu algorithm.
Specifically, for incomplete data, we construct two model selection algorithms that complete in $O(p^2)$ steps: one 
obtains a forest with the maximum posterior probability given the data, 
and the other obtains a forest that converges to the true one as $n$ increases.
We show that the two forests are generally different when some values are missing.
Additionally, we present estimations for benchmark data sets to demonstrate 
that both algorithms work in realistic situations.
Moreover, we derive the conditional entropy provided that no value is missing, and 
we evaluate the per-sample expected redundancy for the universal coding of incomplete data in terms of 
the number of non-missing samples.
\end{abstract}


%


%
%
%
%

\section{INTRODUCTION}
\IEEEPARstart{G}{raphical}
 models have a wide range of applications in various fields, such as signal processing,
coding theory, and bioinformatics \cite{lauritzen}.
Inferring the graphical model from data is a starting point in such applications.

In this paper, we learn a probabilistic relation among $p$ variables $X=(X^{(1)},\cdots,X^{(p)})$
from a data frame that consists of $n$ samples
\begin{eqnarray}
\left.
\begin{array}{ccc}
X^{(1)}=x_{1,1}&\cdots&X^{(p)}=x_{1,p}\nonumber\\
\vdots&\vdots&\vdots\label{eq23}\\
X^{(1)}=x_{n,1}&\cdots&X^{(p)}=x_{n,p}\nonumber\ .
\end{array}
\right\}
n\times p
\end{eqnarray}
The conditional independence relations can generally be expressed by a Bayesian network.
However, estimating the optimal structure given the data frame is difficult
because more than an exponential number of directed acyclic graphs with respect to $p$ are candidates.
In this paper, we assume that the underlying model is a forest rather than a Bayesian network.

The forest learning problem has a long history and has been investigated by several authors.
The basic method was considered by Chow and Liu \cite{c-l}:
connect each pair of variables (vertices) as an edge, as long as no loop is generated by the connection,
in ascending order of mutual information to obtain a forest.
The algorithm assumes that the mutual information values are known, and the resulting tree expresses 
an approximation of the original distribution.
However, we may start from a data frame and connect the edges based on the estimations of mutual information values.
Recently, Liu et al. \cite{liu} estimated its kernel density to prove its consistency for continuous variables 
in a high-dimensional setting, and Tan et al. \cite{tan} restricted the number of edges to prove consistency 
for discrete variables in a different high-dimensional setting.

The approach that we take in this paper is essentially different. We note that by adding an edge to
a forest without creating any loop, the complexity of the forest increases while the likelihood 
of the distribution that the forest expresses improves.
In this sense, the balance should be considered to obtain a correct forest.
In 1993 \cite{uai93}, the author considered a modified estimation of mutual information that takes  the balance  into account
and applied it to the Chow-Liu algorithm.
The resulting undirected graph is not a tree but a forest, and the estimation satisfies consistency, avoiding overfitting
because it minimizes its description length rather than maximizes the likelihood.
Our estimation in this paper is Bayesian and slightly different from \cite{uai93}, although they are essentially equivalent.
Specifically, we find a model that maximizes the posterior probability given the data frame when some of the $pn$ values are missing \cite{roderick,karthika}.

In general, it is computationally difficult to obtain the Bayes optimal solution in model selection with incomplete data.
In fact, suppose that the $p$ variables are binary and that 
 $m$ of the $pn$ values are missing. Thus, we will need to obtain $2^m$ Bayes scores
for each candidate model, where the score is defined by the prior probability
of a model multiplied by the conditional probability of the data frame given the model, because
the $m$ missing values should be marginalized.

An alternative approach to both reduce the computational effort and
obtain a correct model as 
the sample size $n$ increases is to select a model based on the samples such that all the 
$p$ values  are available.
This method ensures consistency, i.e., a correct model is obtained for large $n$
if the size of such samples also becomes large and if an appropriate model selection method is applied to
those samples. However, this method excludes samples such that at least one value is missing,
and it eventually fails to obtain the Bayes optimal model.
In this paper, we assume that the underlying model is a forest rather than a general graphical model
to solve such problems.

The remainder of this paper is organized as follows.
In Section 2, 
assuming that no value is missing in a given data frame,
we consider a Bayes optimal mutual information estimator to avoid such overfitting.
In Section 3, we construct a model selection procedure that maximizes the posterior probability given a data frame
that may contain missing values.
The computation is at most $O(p^2)$ for $p$ variables.
A surprising result is that the model that maximizes the posterior probability does not necessarily converge to
 a correct model as $n$ increases for an incomplete data frame.
Moreover, we illustrate the theoretical results by showing experiments using the Alarm \cite{alarm} and Insurance \cite{insurance} data sets as benchmarks.
In Section 4, we evaluate the code length of each data frame and the expected redundancy per sample 
when some values may be missing, where redundancy is defined by the difference between the expected compression ratio and 
entropy for a given pair of coding and source.
Section 5 summarizes the results and presents directions for future work.

\section{Forest Learning from Complete Data}

We assume that each random variable takes a finite number of values.
By a forest, we mean an undirected graph without any loops.
If  vertex and edge sets are given by
$V=\{1,\cdots,p\}$ and a subset $E$ of 
$\{\{i,j\}|i,j\in V, i\not=j\}$, respectively, then 
we have a distribution 
in the form 
\begin{equation}\label{eq911}
P_X'(X^{(1)},\cdots,X^{(p)}):=\prod_{i\in V}P(X^{(i)})\prod_{\{i,j\}\in E}\frac{P(X^{(i)},X^{(j)})}{P(X^{(i)})P(X^{(j)})}\ .
\end{equation}
Moreover, if we specify 
the probabilities
$\{P(X^{(i)})\}_{i\in V}$ and 
$\{P(X^{(i)},X^{(j)})\}_{\{i,j\}\in E}$, then
the distribution 
(\ref{eq911}) is uniquely determined.
For example, although the distributions
\begin{eqnarray*}
&&P(X^{(2)})P(X^{(1)}|X^{(2)})P(X^{(3)}|X^{(2)})P(X^{(5)}|X^{(2)})\\
&\cdot&P(X^{(4)}|X^{(3)})P(X^{(6)})P(X^{(7)}|X^{(6)})
\end{eqnarray*}
and 
\begin{eqnarray*}
&&P(X^{(4)})P(X^{(3)}|X^{(4)})P(X^{(2)}|X^{(3)})P(X^{(5)}|X^{(2)})\\
&\cdot&(X^{(1)}|X^{(2)})P(X^{(7)})P(X^{(6)}|X^{(7)})
\end{eqnarray*}
may be expressed by directed acyclic graphs as in Figure \ref{fig111} (a) and Figure \ref{fig111} (b), respectively,
both can be expressed by the distribution
\begin{eqnarray*}
&&P(X^{(1)})P(X^{(2)})P(X^{(3)})P(X^{(4)})P(X^{(5)})P(X^{(6)})\\
&\cdot& P(X^{(7)})\cdot \frac{P(X^{(1)},X^{(2)})}{P(X^{(1)})P(X^{(2)})}\cdot \frac{P(X^{(2)},X^{(3)})}{P(X^{(2)})P(X^{(3)})}\\
&\cdot& \frac{P(X^{(2)},X^{(5)})}{P(X^{(2)})P(X^{(5)})}\cdot \frac{P(X^{(3)},X^{(4)})}{P(X^{(3)})P(X^{(4)})}\cdot 
\frac{P(X^{(6)},X^{(7)})}{P(X^{(6)})P(X^{(7)})}
\end{eqnarray*}
and the undirected graph in Figure \ref{fig111} (c), where the vertices $i=1,2,3,4,5,6,7$ and edges $\{j,k\}=\{1,2\}$, $\{2,3\}$, $\{2,5\}$, $\{3,4\}$, and $\{6,7\}$
correspond to $P(X^{(i)})$ and $\displaystyle \frac{P(X^{(j)},X^{(k)})}{P(X^{(j)})P(X^{(k)})}$, respectively.

First, suppose that the distribution $P_X(X^{(1)},\cdots,X^{(p)})$ is known.
We consider maximizing the
Kullback-Leibler divergence $D(P_X||P_X')$ due to 
approximating the true distribution $P_X$ to a distribution $P_X'$ in the form (\ref{eq911}):
\begin{eqnarray}
&&D(P_X||P_X')\nonumber\\
&=&-H(1,\cdots,p)+\sum_{i\in V}H(i)-\sum_{\{i,j\}\in E}I(i,j)\label{eq31}\ ,
\end{eqnarray}
where $H(1,\cdots,p)$, 
$H(i)$, and 
$I(i,j)$
are the entropies of 
$(X^{(1)},\cdots,X^{(p)})$ and
$X^{(i)}$, and the mutual information of ($X^{(i)},X^{(j)}$), respectively.
To minimize $D(P_X||P_X')$,
because the first two terms in (\ref{eq31}) are constants that do not depend on $E$,
we find that maximizing the mutual information sum
$\sum_{\{i,j\}\in E}I(i,j)$ is sufficient. For this purpose, we apply Kruskal's algorithm that,
given symmetric non-negative weights $w(i,j)=w(j,i)$
$i,j\in V$, $i\not=j$, obtains a spanning tree (a connected forest) such that the weight sum is maximized:
let $E$ be the empty set and $E':=\{\{i,j\}|i,j\in V, i\not=j\}$ at the beginning, and 
continue to 
\begin{enumerate}
\item add a  pair $\{i,j\}$ to $E$ with the largest $w(i,j)>0$ among $E'$
if connecting them does not cause any loops to be generated, and
\item remove the $\{i,j\}$ from $E'$ (irrespective of whether the $\{i,j\}$ is connected)
\end{enumerate}
until $E'$ is empty. The Chow-Liu algorithm (1968) \cite{c-l} uses $I(i,j)$ as the weight $w(i,j)$ to minimize 
$D(P_X||P_X')$.
For example, in Figure \ref{figure313}, if
$I(1,2)>I(1,3)>I(2,3)>I(1,4)>I(3,4)>I(2,4)>0$, then $\{1,2\}$ and $\{1,3\}$ are connected at the beginning, but 
$\{2,3\}$ will not be connected because connecting 2 and 3 causes a loop to be generated although the pair has the third largest 
mutual information value.
Furthermore, $\{1,4\}$ is to be connected because it has the fourth largest mutual information value and no loop will be generated.
The procedure terminates at this point because a loop will be generated if any additional pair of vertices is connected.

\begin{figure*}
\begin{center}
\begin{picture}(270,70)(30,0)
\setlength{\unitlength}{0.50mm}

\put(-10,35){(a)}
\put(90,35){(b)}
\put(190,35){(c)}

\small
\put(200,0){\circle{10}}
\put(220,0){\circle{10}}
\put(220,20){\circle{10}}
\put(240,0){\circle{10}}
\put(260,0){\circle{10}}
\put(240,20){\circle{10}}
\put(260,20){\circle{10}}

\put(205,0){\line(1,0){10}}
\put(225,0){\line(1,0){10}}
\put(245,0){\line(1,0){10}}
\put(220,5){\line(0,1){10}}
\put(245,20){\line(1,0){10}}

\put(198,-2){1}
\put(218,-2){2}
\put(238,-2){3}
\put(258,-2){4}
\put(218,18){5}
\put(238,18){6}
\put(258,18){7}

\put(0,0){\circle{10}}
\put(20,0){\circle{10}}
\put(20,20){\circle{10}}
\put(40,0){\circle{10}}
\put(60,0){\circle{10}}
\put(40,20){\circle{10}}
\put(60,20){\circle{10}}

\put(15,0){\vector(-1,0){10}}
\put(25,0){\vector(1,0){10}}
\put(45,0){\vector(1,0){10}}
\put(20,5){\vector(0,1){10}}
\put(45,20){\vector(1,0){10}}

\put(-2,-2){1}
\put(18,-2){2}
\put(38,-2){3}
\put(58,-2){4}
\put(18,18){5}
\put(38,18){6}
\put(58,18){7}

\put(100,0){\circle{10}}
\put(120,0){\circle{10}}
\put(120,20){\circle{10}}
\put(140,0){\circle{10}}
\put(160,0){\circle{10}}
\put(140,20){\circle{10}}
\put(160,20){\circle{10}}

\put(115,0){\vector(-1,0){10}}
\put(135,0){\vector(-1,0){10}}
\put(155,0){\vector(-1,0){10}}
\put(120,5){\vector(0,1){10}}
\put(155,20){\vector(-1,0){10}}

\put(98,-2){1}
\put(118,-2){2}
\put(138,-2){3}
\put(158,-2){4}
\put(118,18){5}
\put(138,18){6}
\put(158,18){7}

\end{picture}
\end{center}
\caption{\label{fig111} Factorizations and their directed and undirected graphs.}
\end{figure*}
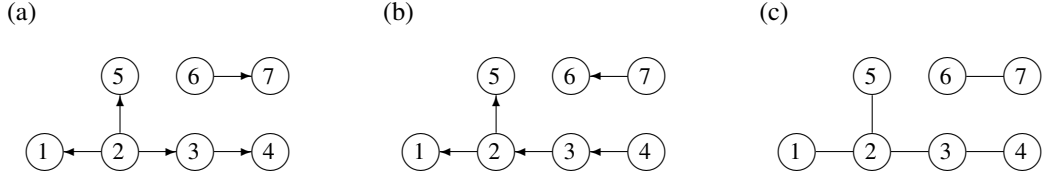

\begin{figure}
\begin{center}
\begin{picture}(210,40)(0,0)
\setlength{\unitlength}{0.40mm}
\put(-3,-2){2}
\put(0,0){\circle{10}}
\put(-3,18){1}
\put(0,20){\circle{10}}
\put(17,18){3}
\put(20,20){\circle{10}}
\put(17,-2){4}
\put(20,0){\circle{10}}

\put(30,10){\vector(1,0){20}}
\put(90,10){\vector(1,0){20}}
\put(150,10){\vector(1,0){20}}

\put(57,-2){2}
\put(60,0){\circle{10}}
\put(57,18){1}
\put(60,20){\circle{10}}
\put(77,18){3}
\put(80,20){\circle{10}}
\put(77,-2){4}
\put(80,0){\circle{10}}
\put(60,5){\line(0,1){10}}

\put(117,-2){2}
\put(120,0){\circle{10}}
\put(117,18){1}
\put(120,20){\circle{10}}
\put(137,18){3}
\put(140,20){\circle{10}}
\put(137,-2){4}
\put(140,0){\circle{10}}
\put(120,5){\line(0,1){10}}
\put(125,20){\line(1,0){10}}

\put(177,-2){2}
\put(180,0){\circle{10}}
\put(177,18){1}
\put(180,20){\circle{10}}
\put(197,18){3}
\put(200,20){\circle{10}}
\put(197,-2){4}
\put(200,0){\circle{10}}
\put(180,5){\line(0,1){10}}
\put(185,20){\line(1,0){10}}
\put(183,17){\line(1,-1){13}}
\end{picture}
\end{center}
\caption{The Chow-Liu algorithm for $I(1,2)>I(1,3)>I(2,3)>I(1,4)>I(3,4)>I(2,4)>0$ \label{figure313}}
\end{figure}
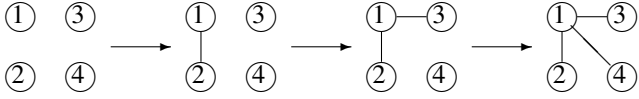

Next, we consider the case in which no distribution but only a data frame is given. In this section, we assume that no value is missing in the data frame.

A naive approach to generate a forest is to estimate a mutual information value $I(i,j)$ by the quantity
\begin{equation}\label{eq96}
I^n(i,j):=\sum_x\sum_y \frac{c(x,y)}{n}\log\frac{c(x,y)/n}{c(x)/n\cdot c(y)/n}
\end{equation}
from the occurrences $c(x),c(y),c(x,y)$
of $X^{(i)},X^{(j)},(X^{(i)},X^{(j)})$ in the $n$ samples and plug $\{I^n(i,j)\}_{i\not=j}$ into
the Chow-Liu algorithm.
Although $I^n(i,j)$ converges to $I(i,j)$ as $n$ increases, $I^n(i,j)$
is always positive; thus,
the Chow-Liu algorithm always generates a spanning tree. For example, two variables are to
be connected for $p=2$ even if they are independent.
This is because maximum likelihood may overfit and in such cases,
eventually no consistency is obtained.


In 1993, Suzuki \cite{uai93} proposed replacing the quantity 
$I^n(i,j)$
by another estimation of mutual information
\begin{equation}\label{eq45}
I^n(i,j)-\frac{1}{2n}(\alpha(i)-1)(\alpha(j)-1)\log n\ ,
\end{equation}
where $\alpha(i)$ and $\alpha(j)$ are the numbers of values that $X^{(i)}$ and $X^{(j)}$ take.
Later, the same author \cite{suzuki12} found that the value of (\ref{eq45}) coincides with
\begin{equation}\label{eq97}
J^n(i,j):=\frac{1}{n}\log \frac{Q^n(i,j)}{Q^n(i)Q^n(j)}
\end{equation}
up to $O(1/n)$ terms, where the quantity $Q^n$, which is termed a Bayes measure and is defined later in this section, 
is computed from $n$ samples with respect to ($X^{(i)},X^{(j)}$) and 
satisfies 
$$0\leq Q^n(i),Q^n(j), Q^n(i,j) \leq 1$$
 and 
 $$\sum Q^n(i), \sum Q^n(j),\sum Q^n(i,j)\leq 1$$
for $n=1,2,\cdots$. 

This method is similar to $I^n(i,j)$ in the sense that $J^n(i,j)\rightarrow I(i,j)$ as $n\rightarrow \infty$,
but it also has a property that it takes a negative value for large $n$ if and only if $X^{(i)}$ and $X^{(j)}$ are independent, written as $X^{(i)}\ci X^{(j)}$.
Moreover, if the prior probability of $X^{(i)}\ci X^{(j)}$ is 
$0<q<1$, then deciding $X^{(i)}\ci X^{(j)}$ if and only if
$$qQ^n(i)Q^n(j)\geq (1-q)Q^n(i,j)$$
maximizes the posterior probability of the decision, which is equivalent to
\begin{equation}\label{eq47}
J^n(i,j)\leq 0 \Longleftrightarrow I(i,j)=0
\end{equation}
when $q=0.5$.
We can show that the decision (\ref{eq47}) is correct for large $n$, which can be stated as follows:
\begin{prop}[Suzuki \cite{suzuki12}]\rm
The decision (\ref{eq47}) is true with probability one for large $n$.
\end{prop}

\begin{figure*}
\begin{center}
\includegraphics[bb=0 0 1000 500, height=8.0cm, width=12.0cm]{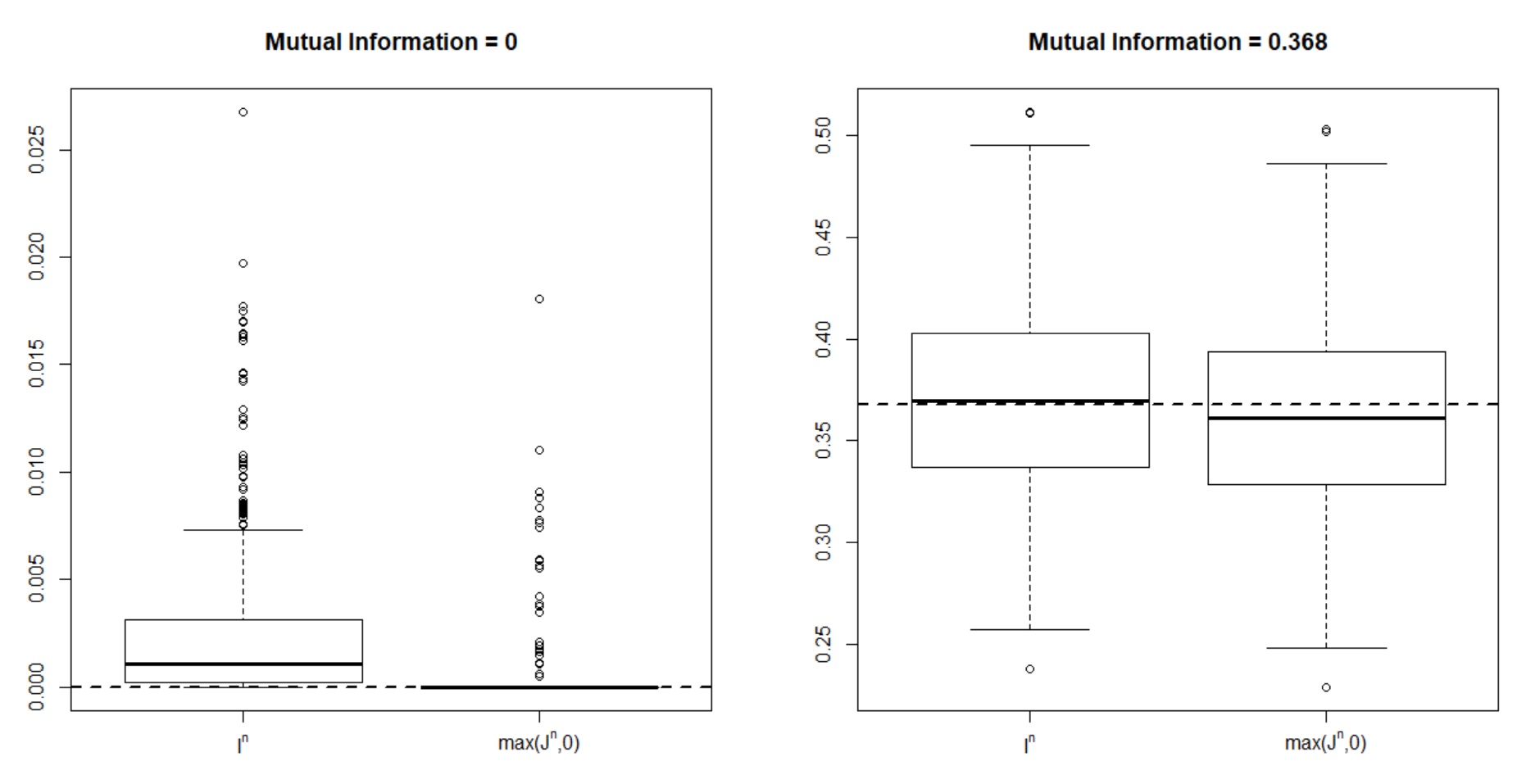}
\end{center}
\caption{The values of $I^n$ and $\max\{J^n,0\}$  when 
the mutual information values are zero (top) and positive (bottom). 
Five-hundred pairs of binary sequences of length 200 are generated such that $P(X=1)=P(Y=1)=0.5$.
The true mutual information values are zero and 0.368 nats with $P(X\not=Y)=0.5$ and $P(X\not=Y)=0.1$, respectively.
\label{fig021}}
\end{figure*}

We shall now define a Bayes measure.
For random variable $X$ that takes zero and one,
let 
$\theta$ be the probability of $X=1$.
Then, the probability that independent sequence $X^n=x^n$ with $c$ ones and $n-c$ zeros occurs can be written as
$\theta^c(1-\theta)^{n-c}$.
If the prior density  $w(\theta)$ of $0\leq \theta\leq 1$ is available, then by integrating
$\theta^c(1-\theta)^{n-c}w(\theta)$ over $0\leq \theta \leq 1$,
we can compute the measure  of $X^n=x^n$ without assuming  any specific 
$\theta$:
$$Q^n(X)=\int_0^1 \theta^c(1-\theta)^{n-c}w(\theta)d\theta\ .$$
If the weight is in the form
$$w(\theta)\propto \theta^{a-1}(1-\theta)^{b-1}$$
with $a,b>0$, then we obtain the Bayes measure 
$Q^n(X)$
as
\begin{equation}\label{eq710}
Q^n(X)=\frac{\Gamma(a+b)}{\Gamma(n+a+b)}\cdot \frac{\Gamma(c+a)}{\Gamma(a)}\cdot \frac{\Gamma(n-c+b)}{\Gamma(b)}\ ,
\end{equation}
where $\Gamma(\cdot)$ is the Gamma function $\Gamma(z)=\int_0^\infty t^{z-1}e^{-t}dt$.
For example, suppose that $a=b=0.5$ and $n=5$. Then, we have
\begin{eqnarray*}
Q^n(X)&=&\frac{1}{5!}
\cdot\underbrace{(c-\frac{1}{2})(c-\frac{3}{2})\cdots \frac{1}{2}}_c\\
&&\cdot\underbrace{(n-c-\frac{1}{2})(n-c-\frac{3}{2})\cdots \frac{1}{2}}_{5-c}\\
&=&
\left\{
\begin{array}{ll}
63/2^8,&c=0,5\\
7/2^8,&c=1,4\\
3/2^8,&c=2,3
\end{array}
\right. \ .
\end{eqnarray*}
We define quantities $Q^n(i),Q^n(j),Q^n(i,j)$ 
similarly to $Q^n(X)$
assuming that $X^{(i)}$, $X^{(j)}$, and $(X^{(i)},X^{(j)})$ take values in $A:=\{0,1,\cdots,\alpha-1\}$ ($\alpha\geq 2$),
$B:=\{0,1,\cdots,\beta-1\}$ ($\beta\geq 2$), and $A\times B$, respectively; the computation for $\{0,1\}^n$ can be extended
to those for $A^n$, $B^n$, and $(A\times B)^n$, respectively.
For example, for $A=\{0,1,\cdots,\alpha-1\}$ with $\alpha\not=2$,
constants $a,b$ and occurrences 
$c,n-c$ are replaced by $a(x)$ and $c(x)$, respectively, for $x=0,1,\cdots,\alpha-1$; thus, the extended formula can be expressed by \cite{causality,suzuki177}
$$Q^n(X)=\frac{\Gamma(\sum_x a(x))}{\Gamma(\sum_{x}(c(x)+a(x)))}\prod_{x=0}^{\alpha-1}\frac{\Gamma(c(x)+a(x))}{\Gamma(a(x))}\ .$$

For completeness, we show the proofs of the derivations from (5) to (4) and Proposition 1 in Appendices A and B, respectively.

\begin{rei}[Experiment]\rm
We show  box plots that depict the 
realizations of $I^n$ and $\max\{J^n,0\}$  when 
the mutual information values are zero  and positive,
and we find that $I^n$ is always larger than $\max\{J^n,0\}$, which is due to its overfitting (Figure \ref{fig021}).
Specifically, $I^n$ cannot detect independence because the value always exceeds zero.
\end{rei}

In contrast, Kruskal's algorithm works even when some weights $w(i,j)$ are either zero or negative: a pair is connected only if
the weight is positive.
If we apply mutual information values  based on (\ref{eq97}) rather than (\ref{eq96})  to the
Chow-Liu algorithm, then it is possible that the value of (\ref{eq97}) is negative, which means that the two variables are independent:
\begin{prop}\rm
A  pair  of nodes need not be connected
even when connecting them does not cause any loops to be generated.
\end{prop}
{\it Proof of Proposition 2}: Kruskal's algorithm connects as an edge only nodes with a positive weight.
\QEDB

Hence, the Chow-Liu algorithm based on (\ref{eq97}) may terminate before
causing overfitting.
Moreover, 
via (\ref{eq911}), 
sequentially choosing an edge with the maximum (\ref{eq97})
is equivalent to choosing a forest  $(V,E)$ with the 
maximum
\begin{equation}\label{eq89}
R^n(E):=\prod_{i\in V}Q^n(i)\prod_{\{i,j\}\in E}\frac{Q^n(i,j)}{Q^n(i)Q^n(j)}\ .
\end{equation}
In fact, the first term on the right-hand side of
\begin{equation}\label{eq75}
-\frac{1}{n}\log R^n(E)=\sum_{i\in V}-\frac{1}{n}\log Q^n(i)-\sum_{\{i,j\}\in E}J^n(i,j)
\end{equation}
is constant irrespective of 
$E$, and minimizing $-\frac{1}{n}\log R^n(E)$ and maximizing the sum of $J^n(i,j)$ values over $\{i,j\}\in E$ are equivalent.
Therefore, if we prepare the uniform prior over the forests, then the Chow-Liu algorithm based on (\ref{eq97})
chooses a forest with the maximum posterior probability given the samples.
If the prior probability is given by 
$$P(E)=K\prod_{\{i,j\}\in E}\frac{1-q(i,j)}{q(i,j)}\ ,$$
where $\displaystyle K:=[\sum_E\prod_{\{i,j\}\in E}\frac{1-q(i,j)}{q(i,j)}]^{-1}$ and $0<q(i,j)<1$, 
then we can add $-\frac{1}{n}\log P(E)$ to (\ref{eq75}) and replace (\ref{eq97}) with \cite{suzuki12}
$$J^n(i,j):=\frac{1}{n}\log \{ \frac{1-q(i,j)}{q(i,j)}\cdot \frac{Q^n(i,j)}{Q^n(i)Q^n(j)}\}\ .$$

Moreover, consistency also holds.
In fact, because $J^n(i,j)\rightarrow I(i,j)$ as $n\rightarrow \infty$, the orders of
$\{J^n(i,j)\}$ and $\{I(i,j)\}$ asymptotically coincide, and from (\ref{eq47}), 
the timing when the procedure terminates is asymptotically correct.

Comparing the Chow-Liu algorithms based on $I^n(i,j)$
and 
$J^n(i,j)$ that maximize
the likelihood and posterior probability, respectively, we find that
\begin{enumerate}
\item $J^n(i,j)$ does not necessarily generate a spanning tree but a forest.
\item the edge set generated by $J^n(i,j)$ is not necessarily a subset of the spanning tree generated by $I^n(i,j)$.
\end{enumerate}

\section{Forest Learning from Incomplete Data}

We consider an extension of the Chow-Liu algorithm based on (\ref{eq97}) such that it
addresses a data frame  that contains missing values.
Specifically, we construct quantities $J^n(i,j)$ and $K^n(i,j)$ to generalize (\ref{eq97}) to respectively obtain 
\begin{enumerate}
\item a forest that maximizes the posterior probability given a data frame
\item a forest that converges to the true one as $n$ increases.
\end{enumerate}

Given a data frame, for each $\{i,j\}$, we compute $\{J^n(i,j)\}$ and $\{K^n(i,j)\}$ based on the samples 
such that none of the values of $X^{(i)}$ and $X^{(j)}$ are missing.

First, we arbitrarily choose a root $r \in V$ for each connected subgraph of the forest.
Let 
$$[r]:=\{k \in \{1,\cdots,n\}| x_{k,r}\ {\rm is\ not\ missing}\}\ ,$$
and $Q^n(r)$ be the Bayes measure with respect to the root $r$ for non-missing values $\{x_{k,r}\}_{k\in [r]}$.
In particular, $r$ represents one of the $p$ variables, and $Q^n(r)$ is obtained via
$$Q^n(r)=\frac{\Gamma(\sum_xa(x))}{\sum_x(c^*(x)+a(x))}\prod_x\frac{c^*(x)+a(x)}{\Gamma(a(x))}\ ,$$
where $c^*(x)$ is the number of the occurrencies of $X^{(r)}=x$ in  $\{x_{k,r}\}_{k\in [r]}$, and 
$x$ ranges over the values that $X^{(r)}$ takes. Note that $c(x)$ with $\sum_{x}c(x)=n$ is replace by $c^*(x)$ with $\sum_{x}c^*(x)\leq n$
because some values are miising.

Because each connected subgraph is a spanning tree, a directed path from its root to each vertex in the subgraph is unique,
and a directed edge set $\vec{E}$ is determined from ${E}$.
We arbitrarily fix a directed edge $(i,j)$ from the upper  $i$ to the lower  $j$, and let
$$[i]:=\{k \in \{1,\cdots,n\}| x_{k,i}\ {\rm is\ not\ missing}\}\ ,$$
$$[j]:=\{k \in \{1,\cdots,n\}| x_{k,j}\ {\rm is\ not\ missing}\}\ ,$$
$$[i,j]:=\{k\in \{1,\cdots,n\}|{\rm neither}\ x_{k,i}\ {\rm nor}\ x_{k,j}\ {\rm are\ missing}\} \ .$$
Moreover, let 
${Q}_{j}^n(i)$, ${{Q}_{i}^n(j)}$,
${{Q}^n(i,j)}$, $Q^n(i)$, and $Q^n(j)$ be the Bayes measures with respect to
$\{x_{k,i}\}_{k\in [i,j]}$,
$\{x_{k,j}\}_{k\in [i,j]}$,
$\{(x_{k,i},x_{k,j})\}_{k\in [i,j]}$,
$\{x_{k,i}\}_{k\in [i]}$, and 
$\{x_{k,j}\}_{k\in [j]}$, respectively.

Suppose that we have a data frame consisting of $p=2$ columns in which 
the values of the two variables are  $(0,*,1,1,*)$ and $(0,1,*,0,*)$, where "*" denotes missing.
Then, $[1,2]=\{1,4\}$, $[1]=\{1,3,4\}$, and $[2]=\{1,2,4\}$ such that 
the scores $Q_2^5(1,2)$, $Q_2^5(1)$, $Q_2^5(2)$, $Q^5(1)$,  and $Q^5(2)$
are associated with the sequences $(00,10)$, $(0,1)$, $(0,0)$, $(0,1,1)$, and $(0,1,0)$.

Since $Q^n(i)$ and $Q^n(j)$ depend on $Q^n_j(i)$ and $Q_i^n(j)$, respectively, rather than on $Q^n(i,j)$,
the Bayes measure with respect to $[i]\cup[j]$ can be evaluated as
$$Q^n(i,j)\cdot \frac{Q^n(i)}{Q^n_j(i)}\cdot \frac{Q^n(j)}{Q^n_i(j)}=Q^n(i)\cdot 
\frac{Q^n(i,j)}{Q^n_{j}(i)}\cdot \frac{Q^n(j)}{Q_{i}^n(j)}\ .
$$
This means that the Bayes measure with respect to the whole forest is evaluated as 
\begin{eqnarray}\label{eq78}
R^n(E)&=&
Q^n(r)
\prod_{(i,j)\in \vec{E}}\{
\frac{Q^n(i,j)}{Q^n_{j}(i)}
\cdot
\frac{Q^n(j)}{Q^n_{i}(j)}
\}\nonumber\\
&=&\prod_{i \in V}Q^n(i)
\prod_{\{i,j\}\in {E}}
\frac{Q^n(i,j)}{Q_{j}^n(i)Q_{i}^n(j)}\ .
\end{eqnarray}
Note that
$\prod_{i \in V}Q^n(i)$ does not depend on the edge set $E$ and that
(\ref{eq89}) is a special case of (\ref{eq78}).

Since maximizing (\ref{eq78}) is equivalent to maximizing the product of $\frac{Q^n(i,j)}{Q^n(i)Q^n(j)}$ in $(i,j)\in E$,
to maximize the posterior probability when the prior distribution over the forests is uniform,
it is sufficient to choose $\{i,j\}$ that maximizes 
\begin{equation}\label{eq8}
{J}^n(i,j):=\frac{1}{n}\log\frac{Q^n(i,j)}{Q_{j}^n(i)Q_{i}^n(j)}
\end{equation}
among the pairs based on the samples 
such that none of the values of $X^{(i)}$ and $X^{(j)}$ are missing.
We find that (\ref{eq8}) contains (\ref{eq97})  as a special case in
which no value is missing 
in the $n$ samples for all the pairs.
We summarize the above discussion as follows:
\begin{teiri}\rm
If we apply $\{{J}^n(i,j)\}_{i\not=j}$ in (\ref{eq8})  to the Chow-Liu algorithm, then we obtain a forest with the maximum posterior probability.
\end{teiri}
{\it Proof of Theorem 1}: Maximizing the Bayes measure (\ref{eq78}) is equivalent to maximizing (\ref{eq8})
at each step of Kruskal's algorithm. 
Since an optimal solution results even with a greedy choice of such pairs, 
we obtain a forest with the maximum posterior probability. 
\QEDB

We find that the value of  (\ref{eq8}) coincides with  the mutual information estimator
\begin{equation}
K^n(i,j)=\frac{1}{n(i,j)}\log\frac{Q^n(i,j)}{Q_{j}^n(i)Q_{i}^n(j)}
\end{equation}
multiplied by the non-missing ratio $n(i,j)/n$, where $n(i,j)$ is the number of non-missing samples for pair $\{i,j\}$,
the cardinality of $[i,j]$.
Therefore, under $n(i,j)\not=n$ for some $i,j=1,\cdots,p$, it is possible that 
maximizing the mutual information estimator $K^n(i,j)$  does not necessarily mean maximizing $J^n(i,j)$. 
Then, we have the following propositions:
\begin{prop}\label{prop2}\rm
Suppose that $n(i,j)\rightarrow \infty$ as
$n\rightarrow \infty$ with probability one
for each $i,j=1,\cdots,p$ ($i\not=j$).
Then,
\begin{equation}
{K}^n(i,j)\longrightarrow I(i,j)
\end{equation}
as $n\rightarrow \infty$ with probability one. 
\end{prop}
\begin{prop}\label{prop3}\rm
Suppose that $n(i,j)\rightarrow \infty$ as
$n\rightarrow \infty$ with probability one
for each $i,j=1,\cdots,p$ ($i\not=j$).
Then,
\begin{equation}
{J}^n(i,j)\leq 0 \Longleftrightarrow I(i,j)= 0 \Longleftrightarrow {K}^n(i,j)\leq 0
\end{equation}
as $n\rightarrow \infty$ with probability one. 
\end{prop}
\begin{prop}\label{prop4}\rm
Suppose that $n(i,j)\rightarrow \infty$ as
$n\rightarrow \infty$ with probability one
for each $i,j=1,\cdots,p$ ($i\not=j$).
Then, if we apply $\{{K}^n(i,j)\}_{i\not=j}$ to the Chow-Liu algorithm,
the generated forest is the true one
as $n\rightarrow \infty$ with probability one. 
\end{prop}
{\it Proofs of Propositions \ref{prop2},\ref{prop3}, and \ref{prop4}}:
If no missing value exists in the original $np$ ones,
from the definitions of $J^n(i,j)$ and $K^n(i,j)$ and
Proposition \ref{prop2}, we have
$$J^n(i,j)=K^n(i,j)\leq 0\Longleftrightarrow I(i,j)=0$$
as $n\rightarrow \infty$ with probability one.
The propositions consider an extended case in which some values may be missing.
Since the occurrence is independent and identically distributed, $K^n(i,j)$ evaluates
independence based on the non-missing $\{(x_{k,h})_{1\leq h\leq p}\}_{k=k_1,\cdots,k_{n(i,j)}}$
with $1\leq k_1\leq \cdots \leq k_{n(i,j)}\leq n$, 
where $[i,j]=\{1,\cdots,k_{n(i,j)}\}$, 
such that 
$K^n(i,j)\rightarrow I(i,j)$ and 
$$K^n(i,j)\leq 0\Longleftrightarrow I(i,j)=0$$
as $n\rightarrow \infty$ with probability one, which proves Proposition \ref{prop2}.
Meanwhile, since $J^n(i,j)=\frac{n(i,j)}{n}K^n(i,j)$, we have 
$$K^n(i,j)\leq 0\Longleftrightarrow J^n(i,j)\leq 0\ ,$$
which proves Proposition \ref{prop3}.
Proposition \ref{prop4} occurs because the orders of $\{K^n(i,j)\}_{i\not=j}$ and $\{I(i,j)\}_{i\not=j}$
asymptotically coincide (Proposition \ref{prop2}) and
the timing when the Chow-Liu algorithm terminates is asymptotically correct
(Proposition \ref{prop3}). \QEDB

Finally, we show that Propositions  \ref{prop2} and \ref{prop4} do not hold for 
$\{J^n(i,j)\}_{i\not=j}$, which means that in model selection with respect to incomplete data,
the model that maximizes the posterior probability 
may not be asymptotically correct.

As an extreme case, if all of the $n$ values are missing for $X^{(3)}$ among
$X^{(1)}, X^{(2)}$, and $X^{(3)}$, then even if 
the values of $I(1,3)$ and $I(2,3)$ are large,
only $\{1,2\}$ will be connected:
\begin{prop}\rm
Maximizing the posterior probability does not imply asymptotic consistency when selecting models with 
incomplete data.
\end{prop}
We prove the proposition by constructing such a case.
In the following example, the $p$ values of $X^{(1)},\cdots,X^{(p)}$ are not missing
with a positive probability:
\begin{rei}\rm
We assume that 
$P(X^{(1)}=1)=P(X^{(2)}=1)=P(X^{(3)}=1)=1/2$, 
$P(X^{(1)}\not=X^{(2)})=P(X^{(1)}\not=X^{(3)})=\epsilon$ with $0<\epsilon<1/2$, 
and $X^{(2)}$ and $X^{(3)}$ are independent.
Then, the true forest should be $E=\{\{1,2\},\{1,3\}\}$
because $P(X^{(2)}\not=X^{(3)})$ is $(1-\epsilon)^2+\epsilon^2>\epsilon$. 
We also find that  
$$K^n(1,2), K^n(1,3)\rightarrow 1-H(\epsilon)$$ and 
$$K^n(2,3) \rightarrow 1-H((1-\epsilon)^2+\epsilon^2)\ .$$
However, 
if we further assume that $X^{(1)}$ is missing with probability 
\begin{equation}\label{eq92}
\frac{H((1-\epsilon)^2+\epsilon^2)-H(\epsilon)}{1-H(\epsilon)}<\delta<1
\end{equation}
and that no values of $X^{(2)}$ and $X^{(3)}$ are missing,
then we find that $\{J^n(i,j)\}$
asymptotically chooses an incorrect forest because
$$J^n(1,2), J^n(1,3)\rightarrow (1-\delta)(1-H(\epsilon))\ ,$$
$$J^n(2,3)\rightarrow 1-H((1-\epsilon)^2+\epsilon^2)\ ,$$
and (\ref{eq92}) is equivalent to 
$(1-\delta)(1-H(\epsilon))<1-H((1-\epsilon)^2+\epsilon^2)$.
\end{rei}

Both Chow-Liu algorithms based on $J^n(,)$ and $K^n(,)$ complete in $O(p^2)$ time.

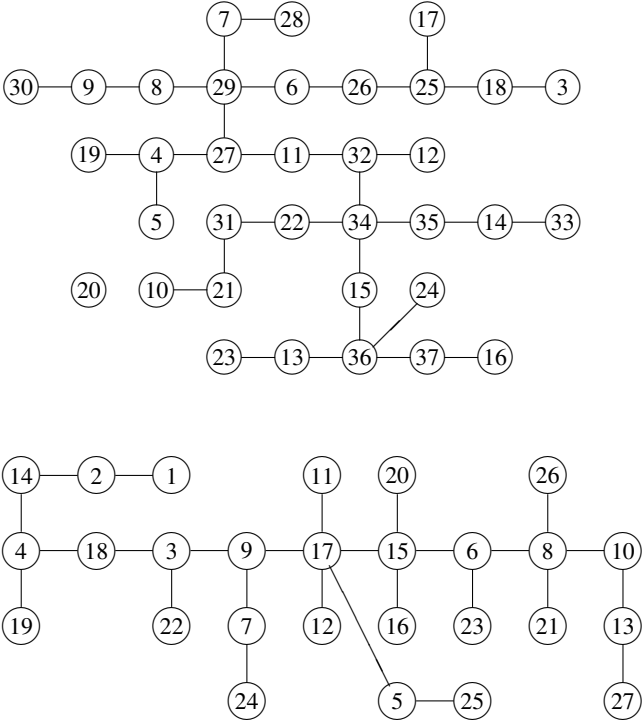
\begin{figure}
\small 
\begin{center}
 \begin{picture}(160,140)(30,0)
 \setlength\unitlength{0.45mm}
  \put(0,80){\circle{10}}
  \put(-5,75){\makebox(10,10)[c]{30}}
  \put(20,80){\circle{10}}
  \put(15,75){\makebox(10,10)[c]{9}}
  \put(40,80){\circle{10}}
  \put(35,75){\makebox(10,10)[c]{8}}
  \put(60,80){\circle{10}}
  \put(55,75){\makebox(10,10)[c]{29}}
  \put(80,80){\circle{10}}
  \put(75,75){\makebox(10,10)[c]{6}}
  \put(100,80){\circle{10}}
  \put(95,75){\makebox(10,10)[c]{26}}
  \put(120,80){\circle{10}}
  \put(115,75){\makebox(10,10)[c]{25}}
  \put(140,80){\circle{10}}
  \put(135,75){\makebox(10,10)[c]{18}}
  \put(160,80){\circle{10}}
  \put(155,75){\makebox(10,10)[c]{3}}
  \put(5,80){\line(1,0){10}}
  \put(25,80){\line(1,0){10}}
  \put(45,80){\line(1,0){10}}
  \put(65,80){\line(1,0){10}}
  \put(85,80){\line(1,0){10}}
  \put(105,80){\line(1,0){10}}
  \put(125,80){\line(1,0){10}}
  \put(145,80){\line(1,0){10}}

  \put(20,60){\circle{10}}
  \put(15,55){\makebox(10,10)[c]{19}}
  \put(40,60){\circle{10}}
  \put(35,55){\makebox(10,10)[c]{4}}
  \put(60,60){\circle{10}}
  \put(55,55){\makebox(10,10)[c]{27}}
  \put(80,60){\circle{10}}
  \put(75,55){\makebox(10,10)[c]{11}}
  \put(100,60){\circle{10}}
  \put(95,55){\makebox(10,10)[c]{32}}
  \put(120,60){\circle{10}}
  \put(115,55){\makebox(10,10)[c]{12}}
  \put(25,60){\line(1,0){10}}
  \put(45,60){\line(1,0){10}}
  \put(65,60){\line(1,0){10}}
  \put(85,60){\line(1,0){10}}
  \put(105,60){\line(1,0){10}}

  \put(60,40){\circle{10}}
  \put(55,35){\makebox(10,10)[c]{31}}
  \put(60,20){\circle{10}}
  \put(55,15){\makebox(10,10)[c]{21}}

  \put(80,40){\circle{10}}
  \put(75,35){\makebox(10,10)[c]{22}}
  \put(100,40){\circle{10}}
  \put(95,35){\makebox(10,10)[c]{34}}
  \put(120,40){\circle{10}}
  \put(115,35){\makebox(10,10)[c]{35}}
  \put(140,40){\circle{10}}
  \put(135,35){\makebox(10,10)[c]{14}}
  \put(160,40){\circle{10}}
  \put(155,35){\makebox(10,10)[c]{33}}

  \put(65,40){\line(1,0){10}}
  \put(85,40){\line(1,0){10}}
  \put(105,40){\line(1,0){10}}
  \put(125,40){\line(1,0){10}}
  \put(145,40){\line(1,0){10}}

  \put(60,100){\circle{10}}
  \put(55,95){\makebox(10,10)[c]{7}}
  \put(80,100){\circle{10}}
  \put(75,95){\makebox(10,10)[c]{28}}
  \put(60,85){\line(0,1){10}}
  \put(65,100){\line(1,0){10}}
  \put(60,65){\line(0,1){10}}
  \put(120,100){\circle{10}}
  \put(115,95){\makebox(10,10)[c]{17}}
  \put(120,85){\line(0,1){10}}
  \put(100,45){\line(0,1){10}}
  \put(40,20){\circle{10}}
  \put(35,15){\makebox(10,10)[c]{10}}
  \put(60,25){\line(0,1){10}}
  \put(40,45){\line(0,1){10}}
  \put(40,40){\circle{10}}
  \put(35,35){\makebox(10,10)[c]{5}}
  \put(45,20){\line(1,0){10}}
  \put(20,20){\circle{10}}
  \put(15,15){\makebox(10,10)[c]{20}}
  \put(120,20){\circle{10}}
  \put(115,15){\makebox(10,10)[c]{24}}
  \put(100,20){\circle{10}}
  \put(95,15){\makebox(10,10)[c]{15}}
  \put(117,16){\line(-1,-1){13}}

  \put(100,5){\line(0,1){10}}
  \put(100,25){\line(0,1){10}}
  \put(65,0){\line(1,0){10}}
  \put(85,0){\line(1,0){10}}
  \put(105,0){\line(1,0){10}}
  \put(125,0){\line(1,0){10}}

  \put(60,0){\circle{10}}
  \put(55,-5){\makebox(10,10)[c]{23}}
  \put(80,0){\circle{10}}
  \put(75,-5){\makebox(10,10)[c]{13}}
  \put(100,0){\circle{10}}
  \put(95,-5){\makebox(10,10)[c]{36}}
  \put(120,0){\circle{10}}
  \put(115,-5){\makebox(10,10)[c]{37}}
  \put(140,0){\circle{10}}
  \put(135,-5){\makebox(10,10)[c]{16}}
  \end{picture}

 \begin{picture}(160,130)(30,0)
 \setlength\unitlength{0.50mm}
  \put(0,60){\circle{10}}
  \put(-5,55){\makebox(10,10)[c]{14}}
  \put(20,60){\circle{10}}
  \put(15,55){\makebox(10,10)[c]{2}}
  \put(40,60){\circle{10}}
  \put(35,55){\makebox(10,10)[c]{1}}
\put(5,60){\line(1,0){10}}
\put(25,60){\line(1,0){10}}

  \put(80,60){\circle{10}}
  \put(75,55){\makebox(10,10)[c]{11}}
  \put(100,60){\circle{10}}
  \put(95,55){\makebox(10,10)[c]{20}}
  \put(140,60){\circle{10}}
  \put(135,55){\makebox(10,10)[c]{26}}

  \put(60,5){\line(0,1){10}}
  \put(160,5){\line(0,1){10}}
  \put(105,0){\line(1,0){10}}

\put(82,36){\line(1,-2){16}}

  \put(60,0){\circle{10}}
  \put(55,-5){\makebox(10,10)[c]{24}}
  \put(100,0){\circle{10}}
  \put(95,-5){\makebox(10,10)[c]{5}}
  \put(120,0){\circle{10}}
  \put(115,-5){\makebox(10,10)[c]{25}}
  \put(160,0){\circle{10}}
  \put(155,-5){\makebox(10,10)[c]{27}}

  \put(0,40){\circle{10}}
  \put(-5,35){\makebox(10,10)[c]{4}}
  \put(20,40){\circle{10}}
  \put(15,35){\makebox(10,10)[c]{18}}
    \put(40,40){\circle{10}}
  \put(35,35){\makebox(10,10)[c]{3}}
  \put(60,40){\circle{10}}
  \put(55,35){\makebox(10,10)[c]{9}}
  \put(80,40){\circle{10}}
  \put(75,35){\makebox(10,10)[c]{17}}
  \put(100,40){\circle{10}}
  \put(95,35){\makebox(10,10)[c]{15}}
  \put(120,40){\circle{10}}
  \put(115,35){\makebox(10,10)[c]{6}}
  \put(140,40){\circle{10}}
  \put(135,35){\makebox(10,10)[c]{8}}
  \put(160,40){\circle{10}}
  \put(155,35){\makebox(10,10)[c]{10}}

  \put(0,20){\circle{10}}
  \put(-5,15){\makebox(10,10)[c]{19}}
 \put(40,20){\circle{10}}
  \put(35,15){\makebox(10,10)[c]{22}}
  \put(60,20){\circle{10}}
  \put(55,15){\makebox(10,10)[c]{7}}
  \put(80,20){\circle{10}}
  \put(75,15){\makebox(10,10)[c]{12}}
  \put(100,20){\circle{10}}
  \put(95,15){\makebox(10,10)[c]{16}}
  \put(120,20){\circle{10}}
  \put(115,15){\makebox(10,10)[c]{23}}
  \put(140,20){\circle{10}}
  \put(135,15){\makebox(10,10)[c]{21}}
  \put(160,20){\circle{10}}
  \put(155,15){\makebox(10,10)[c]{13}}

  \put(0,45){\line(0,1){10}}
  \put(80,45){\line(0,1){10}}
  \put(100,45){\line(0,1){10}}
  \put(140,45){\line(0,1){10}}

  \put(0,25){\line(0,1){10}}
  \put(40,25){\line(0,1){10}}
  \put(60,25){\line(0,1){10}}
  \put(80,25){\line(0,1){10}}
  \put(100,25){\line(0,1){10}}
  \put(120,25){\line(0,1){10}}
  \put(140,25){\line(0,1){10}}
  \put(160,25){\line(0,1){10}}

  \put(5,40){\line(1,0){10}}
  \put(25,40){\line(1,0){10}}
  \put(45,40){\line(1,0){10}}
  \put(65,40){\line(1,0){10}}
  \put(85,40){\line(1,0){10}}
  \put(105,40){\line(1,0){10}}
  \put(125,40){\line(1,0){10}}
  \put(145,40){\line(1,0){10}}

  \end{picture}

\end{center}
\caption{\label{fig101}The generated forests from the Alarm (top) and Insurance (bottom) data sets.
For the meanings of the variables numbered 1-37 and 1-27, consult references \cite{alarm} and \cite{insurance}, respectively.
}
\end{figure}

\begin{figure}
\small 
\begin{center}
 \begin{picture}(160,140)(20,0)
 \setlength\unitlength{0.45mm}
  \put(0,80){\circle{10}}
  \put(-5,75){\makebox(10,10)[c]{30}}
  \put(20,80){\circle{10}}
  \put(15,75){\makebox(10,10)[c]{9}}
  \put(40,80){\circle{10}}
  \put(35,75){\makebox(10,10)[c]{8}}
  \put(60,80){\circle{10}}
  \put(55,75){\makebox(10,10)[c]{29}}
  \put(80,80){\circle{10}}
  \put(75,75){\makebox(10,10)[c]{6}}
  \put(100,80){\circle{10}}
  \put(95,75){\makebox(10,10)[c]{26}}
  \put(120,80){\circle{10}}
  \put(115,75){\makebox(10,10)[c]{25}}
  \put(140,80){\circle{10}}
  \put(135,75){\makebox(10,10)[c]{18}}
  \put(160,80){\circle{10}}
  \put(155,75){\makebox(10,10)[c]{3}}
  \put(5,80){\line(1,0){10}}
  \put(25,80){\line(1,0){10}}
  \put(45,80){\line(1,0){10}}
  \put(65,80){\line(1,0){10}}
  \put(85,80){\line(1,0){10}}
  \put(105,80){\line(1,0){10}}
  \put(125,80){\line(1,0){10}}
  \put(145,80){\line(1,0){10}}

  \put(20,60){\circle{10}}
  \put(15,55){\makebox(10,10)[c]{19}}
  \put(40,60){\circle{10}}
  \put(35,55){\makebox(10,10)[c]{4}}
  \put(60,60){\circle{10}}
  \put(55,55){\makebox(10,10)[c]{27}}
  \put(80,60){\circle{10}}
  \put(75,55){\makebox(10,10)[c]{11}}
  \put(100,60){\circle{10}}
  \put(95,55){\makebox(10,10)[c]{32}}
  \put(120,60){\circle{10}}
  \put(115,55){\makebox(10,10)[c]{12}}
  \put(25,60){\line(1,0){10}}
  \put(45,60){\line(1,0){10}}
  \put(65,60){\line(1,0){10}}
  \put(85,60){\line(1,0){10}}
  \put(105,60){\line(1,0){10}}

  \put(60,40){\circle{10}}
  \put(55,35){\makebox(10,10)[c]{31}}
  \put(60,20){\circle{10}}
  \put(55,15){\makebox(10,10)[c]{21}}

  \put(80,40){\circle{10}}
  \put(75,35){\makebox(10,10)[c]{22}}
  \put(100,40){\circle{10}}
  \put(95,35){\makebox(10,10)[c]{34}}
  \put(120,40){\circle{10}}
  \put(115,35){\makebox(10,10)[c]{35}}
  \put(140,40){\circle{10}}
  \put(135,35){\makebox(10,10)[c]{14}}
  \put(160,40){\circle{10}}
  \put(155,35){\makebox(10,10)[c]{33}}

  \put(65,40){\line(1,0){10}}
  \put(85,40){\line(1,0){10}}
  \put(105,40){\line(1,0){10}}
  \put(125,40){\line(1,0){10}}
  \put(145,40){\line(1,0){10}}

  \put(60,100){\circle{10}}
  \put(55,95){\makebox(10,10)[c]{7}}
  \put(80,100){\circle{10}}
  \put(75,95){\makebox(10,10)[c]{28}}
  \put(60,85){\line(0,1){10}}
  \put(65,100){\line(1,0){10}}
  \put(60,65){\line(0,1){10}}
  \put(120,100){\circle{10}}
  \put(115,95){\makebox(10,10)[c]{17}}
  \put(120,85){\line(0,1){10}}
  \put(100,45){\line(0,1){10}}

  \put(40,20){\circle{10}}
  \put(35,15){\makebox(10,10)[c]{10}}

  \put(60,25){\line(0,1){10}}
  \put(40,45){\line(0,1){10}}

\put(-10,65){\Large A}
\put(10,5){\Large B}
\put(-10,70){\dashbox{3}(80,20){}}
\put(10,10){\dashbox{3}(40,40){}}

  \put(40,40){\circle{10}}
  \put(35,35){\makebox(10,10)[c]{5}}
  \put(45,20){\line(1,0){10}}

  \put(20,20){\circle{10}}
  \put(15,15){\makebox(10,10)[c]{20}}
  \put(120,20){\circle{10}}
  \put(115,15){\makebox(10,10)[c]{24}}
  \put(100,20){\circle{10}}
  \put(95,15){\makebox(10,10)[c]{15}}
  \put(117,16){\line(-1,-1){13}}

  \put(100,5){\line(0,1){10}}
  \put(100,25){\line(0,1){10}}
  \put(65,0){\line(1,0){10}}
  \put(85,0){\line(1,0){10}}
  \put(105,0){\line(1,0){10}}
  \put(125,0){\line(1,0){10}}

  \put(60,0){\circle{10}}
  \put(55,-5){\makebox(10,10)[c]{23}}
  \put(80,0){\circle{10}}
  \put(75,-5){\makebox(10,10)[c]{13}}
  \put(100,0){\circle{10}}
  \put(95,-5){\makebox(10,10)[c]{36}}
  \put(120,0){\circle{10}}
  \put(115,-5){\makebox(10,10)[c]{37}}
  \put(140,0){\circle{10}}
  \put(135,-5){\makebox(10,10)[c]{16}}
  \end{picture}

\end{center}
\caption{\label{fig102}The generated forests from incomplete data sets (Alarm) for $n=20000$. }
\end{figure}
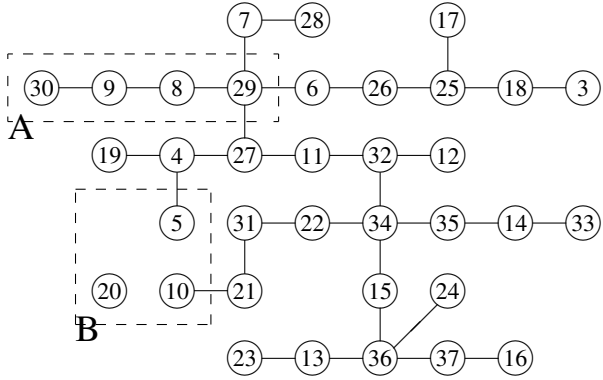

\begin{rei}[Experiments]\rm
For complete data, 
we used the CRAN package BNSL that was developed by Joe Suzuki and Jun Kawahara \cite{bnsl}.
The R package consists of functions that were written using Rcpp \cite{rcpp} and run 50-100 times faster than the usual R functions.
For the experiments, we use the data sets Alarm \cite{alarm} and Insurance \cite{insurance}, 
which are used often as benchmarks for Bayesian network structure learning.
{\small
\begin{verbatim}
library(BNSL); 
mm=mi_matrix(alarm); 
edge.list=kruskal(mm); 
g=graph_from_edgelist(edge.list, directed=FALSE); 
plot(g,vertex.size=1)
\end{verbatim}
}
Before execution, the BNSL package should be installed via 
\begin{verbatim}
install.packages("BNSL")
\end{verbatim}
The functions {\verb+mi_matrix+} and {\verb+kruskal+} obtain the mutual information value matrix and its edge list obtained by Kruskal's algorithm, respectively,
and the last two lines output the graph using the function \verb+plot+ in the {\verb+igraph+} library.
For incomplete data, however, we constructed modified functions that realize $J^n$ and $K^n$ for the experiments.

Figure \ref{fig101} depicts the forests for the complete data with respect to Alarm and Insurance,
which contain $n=20000$ samples for 37 and 27 variables, respectively (consult references \cite{alarm} and \cite{insurance} for the meanings of the variables 
numbered 1-37 and 1-27), 
using the functions in the BNSL package. The forests were generated in a few seconds.

Then, we generated forests with respect to Alarm for the first $n=100, 200, 500, 1000, 2000,$ and $5000$ samples,
but the first ten variables out of 37 were missing with probability $q=0.1, 0.25,$ and $0.50$.
We addressed those data frames  using $J^n$ and $K^n$.
Table \ref{table101} shows the entropy of the generated forests (the data set is random due to the noise, and the resulting forest will
be random). We can consider that the less the entropy, the more stable the estimation,
and we observe that the estimation via $K^n$ is less stable compared with the estimation via $J^n$
for small $n$ and large $p$, which appears to be because the sample size decreases from $n$ to $n(1-q)^2$ on average for the first 10 variables, but
the estimation of $K^n$ is multiplied by $1/(1-q)^2$ on average even if the estimation is based on the small sample size $n(1-q)^2$;
thus, the estimation variance is rather large.
However, for large $n$ and small $q$, the estimation via $K^n$ is more correct than the one via $J^n$.

We expected that for large $n$, the consistent estimation via $K^n$ is closer to the true forest
than that via $J^n$. To examine this expectation, we generated forests using $J^n$ and $K^n$
for $n=20000$ and $q=0.75$, and
the entropy values were 1.798 and 1.195, respectively.
Let $g_1,g_2,g_3,g_4$ be the forests as in Figure \ref{fig102} and Table \ref{tab109}.
The function $K^n$ chooses the correct edges more often than $J^n$. We observe that $g_1$ and $g_2$ are almost close
except for the edges in subgraph $A$. 
Because we added noise to the first 10 variables, it is likely that the mutual information estimation 
between the eighth and ninth variables was underestimated even when $n$ is large.
Function $K^n$ chose $g_1$ and $g_2$ for 97.5\% of the data sets, while $J^n$ chose them for 87 \% of the data sets.
\end{rei}

\begin{table}
\caption{\label{table101} 
The entropies of the forests generated by 
$J^n$ and  $K^n$
for sample sizes $n=100, 200, 500, 1000, 2000,$ and $5000$ and noise probabilities $q=0.1, 0.25,$ and $0.50$. 
}
\begin{center}
\begin{tabular}{l|ll|l|ll|l|ll}
\hline
\multicolumn{3}{c|}{$q=0.1$}&\multicolumn{3}{c|}{$q=0.25$}&\multicolumn{3}{c}{$q=0.50$}\\
\hline
$n$&$J^n$&$K^n$&$n$&$J^n$&$K^n$&$n$&$J^n$&$K^n$\\
\hline
100&5.769&6.392&200&5.569&6.369&200&7.164&7.526\\
200&3.699&4.349&500&3.419&3.823&500&5.751&6.914\\
500&1.963&1.995&1000&2.552&2.513&1000&4.834&5.084\\
1000&0.941&0.840&2000&2.117&2.065&5000&1.366&1.297\\
\hline
\end{tabular}
\end{center}
\end{table}

\begin{table}
\caption{\label{tab109} The forests $g_1,g_2,g_3,g_4$ generated by $J^n$ and $K^n$: $A$ and $B$ are the subgraphs in 
Figure \ref{fig102}, and the edge sets are shown in columns $A$ and $B$. The functions $J^n$ and $K^n$ chose forests $g_1,g_2,g_3,g_4$ 
as in the table. Forest $g_1$ expresses the correct forest in  Figure \ref{fig101} (top).}
\begin{center}
\begin{tabular}{c|c|c|c|c}
\hline
Forests&$A$&$B$&$J^n$&$K^n$\\
\hline
$g_1$&$\{\{8,29\},\{8,9\},\{9,30\}$&$\{\}$&47\%&53.5\%\\
$g_2$&$\{\{9,29\},\{8,9\},\{9,30\}$&$\{\}$&41\%&44.1\%\\
$g_3$&$\{\{8,29\},\{8,9\},\{8,30\}$&$\{\{5,20\}\}$&3.5\%&0.5\%\\
$g_4$&$\{\{8,29\},\{8,9\},\{8,30\}$&$\{\{10,20\}\}$&3\%&0.5\%\\
\hline
\end{tabular}
\end{center}
\end{table}

\section{Universal Coding of Incomplete Data}

In this section, we consider encoding a data frame. 

We claim that the entropy when no value is missing is given by
\begin{equation}\label{eq95}
H(X):=\sum_{i\in V}H(i)-\sum_{\{i,j\}\in E_X}I(i,j)\ ,
\end{equation}
where 
$E_X$ is the edge set obtained by applying the Chow-Liu algorithm to the set $\{I(i,j)\}_{i\not=j}$.
In fact, (\ref{eq95}) is 
$\sum -P_X'\log P_X'$, where $P_X'(X^{(1)},\cdots,X^{(p)})$ is given by (1), and the sum ranges over the values that
$(X^{(1)},\cdots,X^{(p)})$ takes.

Let $H^n(i)$ and $I^n(i,j)$ be the empirical entropy and mutual information of $X^{(i)}$ and $(X^{(i)},X^{(j)})$, respectively.
Then, the description length $L$ based on $\{J^n(i,j)\}_{i\not=j}$ will be
\begin{eqnarray*}
&&\sum_{i\in V}\{nH^n(i)+\frac{\alpha(i)-1}{2}\log n\}\\
&-&\sum_{\{i,j\}\in E_X'}\{nI^n(i,j)-\frac{(\alpha(i)-1)(\alpha(j)-1)}{2}\log n\}+O(1)\ ,
\end{eqnarray*}
where $E_X'$  is the edge set obtained by applying  the Chow-Liu algorithm to the set $\{J^n(i,j)\}_{i\not=j}$.

In information theory, redundancy is defined by the expected description length divided by $n$ minus its entropy.
In this case, it will be at most
\begin{eqnarray*}
&&\frac{EL}{n}-H(X)=\sum_{i\in V}\frac{\alpha(i)-1}{2n}\log n\\
&+&\sum_{\{i,j\}\in E_X}\frac{(\alpha(i)-1)(\alpha(j)-1)}{2n}\log n+O(1/n)\ ,
\end{eqnarray*}
because 
$EH^n(i)=H(i)+O(1/n)$, $EI^n(i,j)=I(i,j)+O(1/n)$, and 
\begin{eqnarray*}
&&E\sum_{\{i,j\}\in E_X'}\{nI^n(i,j)-\frac{(\alpha(i)-1)(\alpha(j)-1)}{2}\log n\}\\
&\geq &n\sum_{\{i,j\}\in E_X}\{EI^n(i,j)-E\frac{(\alpha(i)-1)(\alpha(j)-1)}{2n}\log n\}\\
&= &n\{\sum_{\{i,j\}\in E_X}I(i,j)-\sum_{\{i,j\}\in E_X}\frac{(\alpha(i)-1)(\alpha(j)-1)}{2n}\log n\}\\
&&+O(1)
\end{eqnarray*}

We now consider the general case in which some values are missing.
For
this purpose,
we define a sequence (source) of random variables $Y=(Y^{(1)},\cdots,Y^{(p)})$
such that each 
$Y^{(i)}$ takes either zero or one.
We assume that $Y$ is stationary ergodic, as are
$\{Y^{(i)}\}_{i\in V}$ and  $\{(Y^{(i)},Y^{(j)})\}_{i\not=j}$.
We define $X=(X^{(1)},\cdots,X^{(p)})$ by 
$$
X^{(i)}
\left\{
\begin{array}{ll}
\in\{0,\cdots,\alpha(i)-1\},&Y^{(i)}=1\\
=\alpha(i),&Y^{(i)}=0
\end{array}
\right.\ ,
$$
where $\alpha(i)\geq 2$.
Let
$r(i)$ and $r(i,j)$ be the stationary probabilities of $Y^{(i)}=1$ and $Y^{(i)}=Y^{(j)}=1$ for $i\not=j$, respectively.
Then, we extend
(1)  into
\begin{eqnarray*}
&&P_{X|Y}'(X^{(1)},\cdots,X^{(p)}|Y^{(1)},\cdots,Y^{(p)})\\
&=&\prod_{i\in V}P(X^{(i)})^{Y(i)}\prod_{\{i,j\}\in E_{X|Y}}\{\frac{P(X^{(i)}, X^{(j)}))}{P(X^{(i)}P(X^{(j)}))}\}^{Y(i)Y(j)}\ ,
\end{eqnarray*}
where $E_{X|Y}$ is the edge set obtained by applying the Chow-Liu algorithm to the set $\{r(i,j)I(i,j)\}_{i\not=j}$.
We note that $Y(i)=1$ for $i\in V$ and $Y(i)=Y(j)=1$ for $\{i,j\}\in E$ imply the original probability (1).
Then, the entropy $H(X)$ in  (\ref{eq95}) becomes the conditional entropy  of $X$ given $Y$:
\begin{eqnarray*}
&&H(X|Y)\\
&:=&\sum -P_Y(Y^{(1)},\cdots,Y^{(p)})P_{X|Y}'\log P_{X|Y}'\\
&=&\sum_{i\in V}r(i)\sum -P_{X|Y}(X^{(i)}|Y^{(i)}=1)\\
&&\cdot \log P_{X|Y}(X^{(i)}|Y^{(i)}=1)-\sum_{\{i,j\}\in E_{X|Y}}r(i,j)\\&&\sum P_{X|Y}(X^{(i)},X^{(j)}|Y^{(i)}=Y^{(j)}=1)\\
&&\cdot \log 
\{P_{X|Y}(X^{(i)}, X^{(j)}|Y^{(i)}=Y^{(j)}=1)/\\&&
[P_{X|Y}(X^{(i)}|Y^{(i)}=Y^{(j)}=1)\\&&
\cdot P_{X|Y}(X^{(j)}|Y^{(i)}=Y^{(j)}=1)]\}\\
&=&\sum_{i\in V}r(i)H(i)-\sum_{\{i,j\}\in E_{X|Y}}r(i,j)I(i.j)
\end{eqnarray*}
Moreover, the description length based on $\{J^n(i,j)\}_{i\not=j}$ is at most
\begin{eqnarray*}
&&\sum_{i\in V}\{n(i)H^n(i)+\frac{\alpha(i)-1}{2}\log n(i)\}\\
&&-\sum_{\{i,j\}\in E_{X|Y}'}\{n(i,j)I^n(i,j)\\&&-\frac{(\alpha(i)-1)(\alpha(j)-1)}{2}\log n(i,j)\}+O(1)\ ,
\end{eqnarray*}
where $E_{X|Y}'$  is the edge set obtained by applying  the Chow-Liu algorithm to the set $\{J^n(i,j)\}_{i\not=j}$.

Since 
$E[n(i)/n]=r(i)$, $E[n(i,j)/n]=r(i,j)$, $EH^n(i)=H(i)+O(1/n)$, $EI^n(i,j)=I(i,j)+O(1/n)$, and 
\begin{eqnarray*}
&&E\sum_{\{i,j\}\in E_{X|Y}'}\{n(i,j)I^n(i,j)\\&&-\frac{(\alpha(i)-1)(\alpha(j)-1)}{2}\log n(i,j)\}\\
&\geq &n\sum_{\{i,j\}\in E_{X|Y}}\{E[\frac{n(i,j)}{n}I^n(i,j)]\\&&-E\frac{(\alpha(i)-1)(\alpha(j)-1)}{2n}\log n(i,j)\}\\
&= &n\{\sum_{\{i,j\}\in E_{X|Y}}r(i,j)I(i,j)\\&&-\sum_{\{i,j\}\in E_{X|Y}}\frac{(\alpha(i)-1)(\alpha(j)-1)}{2n}\log n(i,j)\}\\
\end{eqnarray*}
up to $O(1)$ terms, we have a final result:
\begin{teiri}\rm
The redundancy for the general case in which some values are missing is at most
\begin{eqnarray*}
&&\frac{EL}{n}-H(X|Y)=\sum_{i\in V}\frac{\alpha(i)-1}{2n}\log n(i)\\
&-&\sum_{\{i,j\}\in E_{X|Y}}\frac{\{\alpha(i)-1\}\{\alpha(j)-1\}}{2n}\log n(i,j)+O(1/n)\ ,
\end{eqnarray*}
\end{teiri}

\section{Concluding Remarks}

In statistics, how to address missing values is an important issue.
If one wishes to obtain correct dependencies in a data frame with $n$ samples and $p$
variables,
then the true model is obtained as $n$ increases by removing the records that contain at least one missing value.

However,  in general, for large $p$, because such records are few, a large $n$ is required.
To utilize the data instead, 
one might wish to obtain Bayes optimal dependencies.
However, the computation is exponential with the number of missing locations.
The current paper suggested that the best compromise would be to model dependencies
using a forest rather than Bayesian and Markov networks, and it found the following novel insights:
\begin{enumerate}
\item the model that maximizes the posterior probability (minimizes the description length) does not increase
the computation, and
\item it is possible that the estimated Bayes optimal model does not converge to the true one as $n\rightarrow \infty$.
\end{enumerate}

In model selection, we often expect consistency by maximizing the posterior probability.
This paper suggests that such an estimation might be useless if a missing value exists.

As a future work, we will consider exactly when maximizing the posterior probability and consistent estimation 
do not coincide in model selection when some values are missing.

\section*{Appendix A: Proof of (5) from (4)}
We utilize Stirling's formula $\log \Gamma(z)\sim -z+(z-\frac{1}{2})\log z$, where 
$A\sim B$ denotes that $|A-B|$ is bounded by a constant.
From
\begin{eqnarray*}
&&\log\Gamma(n+\frac{\alpha}{2})\sim -(n+\frac{\alpha}{2})+(n+\frac{\alpha-1}{2})\log (n+\frac{\alpha}{2})\\
&\sim &
-n+(n+\frac{\alpha-1}{2})\log n
\end{eqnarray*}
and
$$\log \Gamma(c+\frac{1}{2})\sim -c+c\log c$$
for $0\leq c\leq n$, we have
$$-\log Q^n(i)\sim \sum_{x=0}^{\alpha-1}c(x)\log \frac{n}{c(x)}+\frac{\alpha-1}{2}\log n $$
for $0\leq c(x)\leq n$, $x=0,\cdots,\alpha-1$, $\sum_{x=0}^{\alpha-1}c(x)=n$, and 
$$Q^n(i)=\frac{\Gamma(\alpha/2)}{\Gamma(n+\alpha/2)}\prod_{x=0}^{\alpha-1}\frac{\Gamma(c(x)+1/2)}{\Gamma(1/2)}\ .$$
Similarly, we have 
$$-\log Q^n(i,j)\sim \sum_{x=0}^{\alpha-1}\sum_{y=0}^{\beta-1}c(x,y)\log \frac{n}{c(x,y)}+\frac{\alpha\beta-1}{2}\log n$$
for 
$0\leq c(x,y)\leq n$, $x=0,\cdots,\alpha-1$, $y=0,1,\cdots,\beta-1$, $\sum_{x=0}^{\alpha-1}\sum_{y=0}^{\beta-1}c(x,y)=n$, and 
 $$Q^n(i,j)=\frac{\Gamma(\alpha\beta/2)}{\Gamma(n+\alpha\beta/2)}\prod_{x=0}^{\alpha-1}\prod_{y=0}^{\beta-1}\frac{\Gamma(c(x,y)+1/2)}{\Gamma(1/2)}\ ,$$
which implies that
$$\log \frac{Q^n(i,j)}{Q^n(i)Q^n(j)}\sim nI^n(i,j)-\frac{(\alpha-1)(\beta-1)}{2}\log n\ .$$

\section*{Appendix B: Proof of Proposition 1}
If $X^{(i)}$ and $X^{(j)}$ are not independent, 
then the estimate $I^n(i,j)$
converges to the mutual information $I(i,j)>0$.
On the other hand, $\frac{(\alpha-1)(\beta-1)}{2n}\log n$ converges to zero, which means that $J^n(i,j)>0$
with probability one as $n\rightarrow \infty$.

Suppose that $X^{(i)}$ and $X^{(j)}$ are independent.
Then, it is known \cite{c35} that 
$$2nI^n(i,j)\sim \sum_{x=0}^{\alpha-1}\sum_{y=0}^{\beta-1}Z_{xy}^2$$
with 
$$Z_{xy}:=\frac{c(x,y)-np(x)q(y)}{\sqrt{np(x)q(y)}}\ ,$$
where $p(x)$ and $q(y)$ are the probabilities of $X^{(i)}=x$ and $X^{(j)}=y$, respectively.

Then, it is sufficient to show that
\begin{equation}\label{eq81}
\sum_{x=0}^{\alpha-1}\sum_{y=0}^{\beta-1}Z_{xy}^2\leq 2(1+\epsilon)(\alpha-1)(\beta-1)\log \log n
\end{equation}
for an arbitrarily small $\epsilon>0$ with probability one as $n\rightarrow \infty$
because the right-hand side of (4) is at most $(\alpha-1)(\beta-1)\log n/n$.

For the matrix $Z=[Z_{xy}]\in {\mathbb R}^{\alpha\beta}$, 
$$u_0=[\sqrt{p(0)},\cdots,\sqrt{p(\alpha-1)}]^T\ ,$$ and
$$v_0=[\sqrt{q(0)},\cdots,\sqrt{q(\beta-1)}]^T\ ,$$
we have $u_0^TZ=0$ and $Zv_0=0$.
Let $u_1,\cdots,u_{\alpha-1}$ and $v_1,\cdots,v_{\beta-1}$ be such that 
$U=[u_0,u_1,\cdots,u_{\alpha-1}]$ and $V=[v_0,v_1,\cdots,v_{\beta-1}]$
are orthogonal matrices.
Then, we find that the $\alpha\beta$ square sums of the elements in $R:=U^TZV$ and $Z$ are the same,
and at most $(\alpha-1)(\beta-1)$ values are nonzero in $R=[R_{k,h}]$:
\begin{equation}\label{eq82}
\sum_{x=0}^{\alpha-1}\sum_{y=0}^{\beta-1}Z_{xy}^2=\sum_{k=1}^{\alpha-1}\sum_{h=1}^{\beta-1}R_{k,h}^2
\end{equation}
One can check that $R_{i,j}=\frac{1}{\sqrt{n}} \sum_{r=1}^n {Y_{k,h,r}}$ with 
$$Y_{k,h,r}:=\sum_{x=0}^{\alpha-1}\sum_{y=0}^{\beta-1} u_{k,x}v_{h,y}\frac{I(X_r^{(i)}=x,X_r^{(j)}=y)-p(x)q(y)}{\sqrt{p(x)q(y)}}$$
satisfies $EY_{k,h,r}^2=1$, and $EY_{k,h,r}Y_{k',h',r}=0$ for $(k,h)\not=(k',h')$,
where $I(A)=1$ if event $A$ is true, and $I(A)=0$ otherwise. In fact,
$$E(\frac{I(X_r^{(i)}=x,X_r^{(j)}=y)-p(x)q(y)}{\sqrt{p(x)q(y)}})^2=1-p(x)q(y)$$
\begin{eqnarray*}
&&E[(\frac{I(X_r^{(i)}=x,X_r^{(j)}=y)-p(x)q(y)}{\sqrt{p(x)q(y)}})\\
&&(\frac{I(X_r^{(i)}=x',X_r^{(j)}=y')-p(x')q(y')}{\sqrt{p(x')q(y')}})]\\
&=&-\sqrt{p(x)q(y)p(x')p(y')}
\end{eqnarray*}
for $(x,y)\not=(x',y')$, $\sum_{x=0}^{\alpha-1}u_{k,x}\sqrt{p(x)}=0$, and $\sum_{y=0}^{\beta-1}v_{h,y}\sqrt{q(x)}=0$, such that
\begin{eqnarray*}
&&EY_{k,h,r}^2\\
&=&
\sum_{x=0}^{\alpha-1}\sum_{y=0}^{\beta-1} u_{k,x}^2v_{h,y}^2E(\frac{I(X_r^{(i)}=x,X_r^{(j)}=y)-p(x)q(y)}{\sqrt{p(x)q(y)}})^2
\\&&-\sum\sum\sum\sum u_{k,x}v_{h,y}u_{k,x'}v_{h,y'}\\&&E
[(\frac{I(X_r^{(i)}=x,X_r^{(j)}=y)-p(x)q(y)}{\sqrt{p(x)q(y)}})\\
&&(\frac{I(X_r^{(i)}=x',X_r^{(j)}=y')-p(x')q(y')}{\sqrt{p(x')q(y')}})]\\
&=&
\sum_{x=0}^{\alpha-1}\sum_{y=0}^{\beta-1} u_{k,x}^2v_{h,y}^2(1-p(x)q(y))\\&&-
\sum\sum\sum\sum u_{k,x}v_{h,y}u_{k,x'}v_{h,y'}\sqrt{p(x)p(x')q(y)q(y')}\\
&=&
\sum_{x=0}^{\alpha-1}\sum_{y=0}^{\beta-1} u_{k,x}^2v_{h,y}^2-
\sum_{x=0}^{\alpha-1}u_{k,x}\sqrt{p(x)}\\&&\cdot \sum_{x'=0}^{\alpha-1}u_{k,x'}\sqrt{p(x')}\cdot
\sum_{y=0}^{\beta-1} v_{h,y}\sqrt{q(y)}\cdot \sum_{y=0}^{\beta-1}v_{h,y'}\sqrt{q(y')}\\
&=&1\ ,
\end{eqnarray*}
where the sums $\sum\sum\sum\sum$ range over all $k,k'=0,1,\cdots,\alpha-1$ and $h,h'=0,1,\cdots,\beta-1$ s.t. $(k,h)\not=(k',h')$,
and
\begin{eqnarray*}
&&EY_{k,h,r}Y_{k',h',r}\\
&=&
\sum_{x=0}^{\alpha-1}\sum_{y=0}^{\beta-1}\sum_{x'=0}^{\alpha-1}\sum_{y'=0}^{\beta-1} u_{k,x}v_{h,y}u_{k',x'}v_{h',y'}\\&&
E(\frac{I(X_r^{(i)}=x,X_r^{(j)}=y)-p(x)q(y)}{\sqrt{p(x)q(y)}})\\
&&(\frac{I(X_r^{(i)}=x',X_r^{(j)}=y')-p(x')q(y')}{\sqrt{p(x')q(y')}})\\
&=&
\sum_{x=0}^{\alpha-1}\sum_{y=0}^{\beta-1}\sum_{x'=0}^{\alpha-1}\sum_{y'=0}^{\beta-1}
u_{k,x}v_{h,y}u_{k',x'}v_{h',y'}\sqrt{p(x)q(y)p(x')q(y')}\\
&=&
\sum_{x=0}^{\alpha-1}u_{k,x}\sqrt{p(x)}\cdot \sum_{x'=0}^{\alpha-1}u_{k,x'}\sqrt{p(x')}\cdot
\sum_{y=0}^{\beta-1} v_{h,y}\sqrt{q(y)}\\&&\cdot \sum_{y=0}^{\beta-1}v_{h,y'}\sqrt{q(y')}\\
&=&0\ .
\end{eqnarray*}

We apply the law of the iterated logarithm.
\begin{hodai}[Billingsley\cite{billingsley}]\rm
Let $X_1,\cdots,X_n$ be independent, and each $X_i$ has zero mean  and unit variance.
Then, for any small $\epsilon$, 
$$X_1+\cdots +X_n \leq(1+\epsilon)\sqrt{2n\log \log n}$$
with probability one as $n\rightarrow \infty.$
\end{hodai}
From the lemma, we obtain the following inequality:
\begin{equation}\label{eq83}
\sqrt{n}R_{k,h}=\sum_{r=1}^n {Y_{k,h,r}}\leq (1+\epsilon)\sqrt{2n\log \log n}
\end{equation}
for an arbitrarily small $\epsilon>0$ with probability one as $n\rightarrow \infty$.
From (\ref{eq82}) and (\ref{eq83}), we have (\ref{eq81}), which completes the proof.

\bibliography{2017-7-17}

\end{document}